\documentclass[prd,aps,preprint,preprintnumbers,nofootinbib]{revtex4}
\usepackage{epsfig,footnote}
\usepackage{ulem}
\usepackage{color}
\usepackage{array}
\usepackage{amssymb}
\usepackage{amsmath}
\usepackage{graphicx,subfigure}
\usepackage{longtable}
\usepackage{verbatim}
\usepackage{amsfonts}
\usepackage{hyperref}
\usepackage{cancel}
\newcommand{\degree}{^\circ}
\begin{document}
\title{Analysis of three-body charmed $B$ meson decays under the factorization-assisted
topological-amplitude approach}
\author{Si-Hong Zhou\footnote{shzhou@imu.edu.cn}, Run-Hui Li\footnote{lirh@imu.edu.cn} and Zheng-Yi Wei}
\affiliation{School of Physical Science and Technology,
Inner Mongolia University, Hohhot 010021, China}
\author{Cai-Dian L\"u\footnote{lucd@ihep.ac.cn}}
\affiliation{Institute of High Energy Physics, CAS, Beijing 100049, China }
\affiliation{ School of Physics, University of Chinese Academy of Sciences, Beijing 100049, China
}
\date{\today}
\begin{abstract}
We analyze the quasi-two-body charmed $B$ decays $B^{+,0}_{(s)} \to D_{(s)}^* P_2  \to D_{(s)}  P_1 P_2$
with $P_{1,2}$ as a pion or kaon. The intermediate processes $B_{(s)} \to D_{(s)}^* P_2 $ are calculated
with the factorization-assisted topological-amplitude approach and the resonant effects are calculated with the Breit-Wigner formalism.
Taking all $P$-wave resonance states $ \bar D_{(s)}^*$ into consideration, we present the related branching fractions, calculate the Breit-Wigner-tail effects, and investigate the flavor $SU(3)$ breaking effects. Most of our branching fractions are consistent with the perturbative QCD approach's predictions as well as the current experimental data. With more precision calculation of the intermediate two body charmed $B$ meson decays, our quasi-two-body $B$ decays calculation has significantly less theoretical uncertainty than the perturbative QCD approach. Many of those  channels without any experimental data will be confronted with the future   more accurate experiment measurements. Our results of the Breit-Wigner-tail effects also agree with the experimental data very well. In $B^0$ decays this effect can reach approximately to $5\%$. It is also found that the Breit-Wigner-tail effects are not sensitive to the widths of their corresponding resonances.
\end{abstract}

\maketitle

\section{Introduction}\label{Introduction}
$B$ meson non-leptonic decays     are  very important for the study of many frontier topics,
such as the mechanism of $CP$ violation and the emergence of quantum chromodynamics.
 The three-body hadronic decays of $B$ mesons, with many kinds of intermediate  states   parameterized by the
 sequential two body decays, provide opportunities for the study
 of these common topics and of the hadron spectroscopy. So far, plenty of data on the three-body hadronic $B$ meson
 decays have been collected by BABAR, 
  Belle  
  and LHCb 
   experiments
 using Dalitz plot technique, which induced   a lot of     theoretical studies.

On the theoretical side,
contrary to two-body $B$ decays, which have been analyzed extensively in the past two decades within the
frameworks of the factorization approach, QCD factorization approach \cite{Beneke:2000ry}, perturbative QCD (PQCD) \cite{Lu:2000em}, soft collinear effective theory \cite{Bauer:2000yr}
and some model independent approaches
such as the factorization-assisted topological-amplitude (FAT) approach \cite{Li:2012cfa}, three-body $B$ decays are more complicated. On one hand the three-body decays receive both resonant and nonresonant contributions, on the other hand factorization of these decays has not been completely proved. 
Some theoretical models aimed to calculate the non-resonant effects of some three-body $B$ decays are in development such as the heavy meson chiral perturbation theory  \cite{Cheng:2002qu,Cheng:2007si,Cheng:2013dua},
a model combing the heavy quark effective theory and chiral Lagrangian \cite{Fajfer:2004cx},
and the  perturbative QCD approach by introducing di-meson distribution amplitudes \cite{Chen:2002th}.
  The three-body $B$ meson decays are usually dominated by intermediate resonances. They proceed formally as quasi-two-body decays with
an intermediate sate of a vector or scalar resonance and a ``bachelor"  meson.
The quasi-two-body framework is employed by the PQCD approach
 \cite{Chen:2002th, Wang:2014ira, Wang:2015uea, Wang:2016rlo,Ma:2016csn, plb791-342, Wang:2020plx,Chai:2021kie} and
 some phenomenological analysis based on factorization \cite{Cheng:2002qu,Cheng:2007si,Cheng:2013dua}.
Some three-body $B$ decays are also analyzed by the QCD factorization approach \cite{npb899-247, jhep2006-073, Huber:2020pqb}
and implemented within the $U$-spin, isospin and flavour SU(3) symmetries (see also Refs~\cite{prd72-094031,plb728-579,prd72-075013,prd84-056002,plb726-337,prd89-074043}).

In this work, we concentrate on three-body charmed $B$ decays $B_{(s)} \to D_{(s)} P_1 P_2$,
where $P_{1,2}$ is a pion or kaon, via quasi-two-body decays $B_{(s)} \to D_{(s)}^* P_2  \to D_{(s)}  P_1 P_2$
with $P_1$ stemming from the decay of the resonance and $P_2$ representing a ``bachelor" light meson.
The Belle, BABAR and LHCb Collaborations have achieved brilliant progress in identifying the excited charmd
states and also found their off-shell effects are indispensable
\cite{Kuzmin:2006mw,Aaij:2015vea,Aaij:2015kqa,
Aaij:2015sqa,Aaij:2016fma,Wiechczynski:2014kxh,prd90-072003}.
The off-shell effect, also called the Breit-Wigner-tail (BWT) effect,
is the contribution when the pole mass of $D_{(s)}^* $ is smaller than the invariant mass of $D_{(s)}  P_1$
or, in the other case, the contribution with the on-shell effect excluded by a cut on the invariant mass of $D_{(s)}  P_1$.
Recently, motivated by the experimental measurements
on the off-shell effects of $D^*$ meson in  $B \to D \pi \pi (K)$ decays, theoretical attention is paid to these virtual effects~\cite{prd99-073010,plb791-342}.
A new systematic research on the three-body charmed $B$ meson decays
including $D_{(s)}^* $ resonant states have been done subsequently in the PQCD approach~\cite{Chai:2021kie}.
About $5\%$ off-shell contributions of $D^*$ resonances and tiny contributions ($<1 \%$) from $D_{s}^* $ resonances are found.

The theoretical uncertainty of the two-step process of $B_{(s)} \to D_{(s)}^* P_2  \to D_{(s)}  P_1 P_2$
is dominated by the uncertainty from two-body non-leptonic $B$ decays $B_{(s)} \to D_{(s)}^* P$. Large non-factorizable contribution  has been found in the  factorization approach~\cite{zpc34-103,plb318-549,ijmpa24-5845} and the PQCD approach~\cite{prd67-054028, prd69-094018}. Since the charm quark mass scale is also involved in these decays in addition to the $b$ quark mass scale, the power corrections $m_c/m_b$ of these decays are very difficult to calculate.
In a previous work~\cite{Zhou:2015jba}, two of us (S-H Z, C-D L), with other colleagues, utilize  the framework of conventional topological diagram approach to group  the decay amplitudes by different electroweak Feynman diagrams\cite{Cheng:2014rfa}. A global fit with all experimental data of
 these decays to extract the topological amplitudes   including the nonfactorizable QCD contributions as well as the $SU(3)$ breaking effects is performed.
 This so-called FAT approach
 \cite{Zhou:2015jba,Li:2012cfa,Li:2013xsa,Zhou:2016jkv,Jiang:2017zwr,Zhou:2019crd} gives the most precise decay amplitudes of the
two body $B$ meson decays with charmed meson final states. For example,
the amplitudes of color-suppresses topological diagram ($C$) of Fig.~\ref{TCE}(b),
dominated by the nonfactorizable QCD effect is larger in the FAT approach than that in other approaches.
In this paper we apply the FAT approach to quasi-two-body $B$ meson decays, which will significantly reduce the theoretical uncertainty.  The Breit-Wigner formalism is used to describe the resonance and a strong coupling accounts for the subsequent decay $D_{(s)}^*\to D_{(s)} P_1$ .

This paper is organized as follows. In Section \ref{sec:2}, the theoretical framework is introduced. The numerical results and
discussions are collected in Section \ref{sec:3}. Section \ref{sec:4} is a summary.

\section{FACTORIZATION OF AMPLITUDES FOR TOPOLOGICAL DIAGRAMS}\label{sec:2}

\begin{figure}
\begin{center}
\scalebox{0.5}{\epsfig{figure=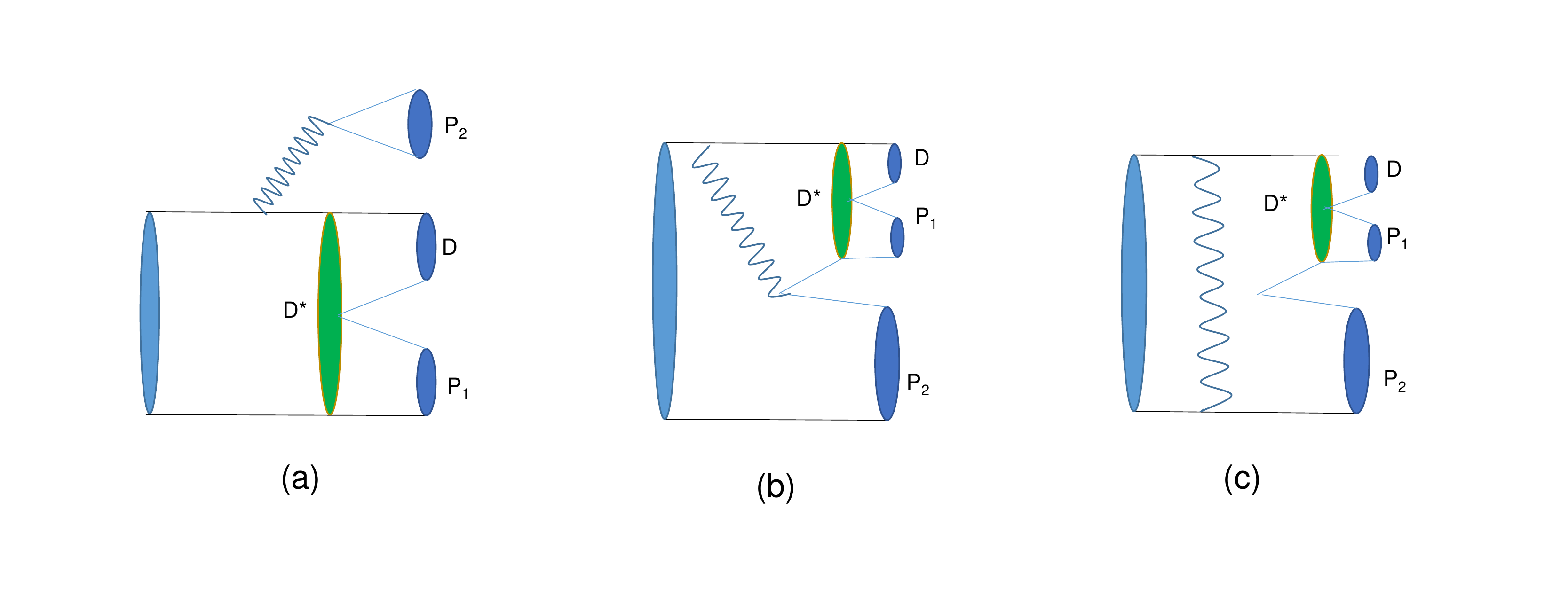}}
\vspace{-1.7cm}
\caption{Topological diagrams of $B_{(s)} \to D_{(s)}^* P_2  \to D_{(s)}  P_1 P_2$ under the framework of quasi-two-body decay with the wave line representing a $W$ boson:
  (a) the color-favored tree diagram $T$;
  (b) the color-suppressed tree diagram $C$;
  (c) the W-exchange  diagram $E$.}
\label{TCE}
\end{center}
\end{figure}

Under the framework of quasi-two-body decay, the $B_{(s)}  \to D_{(s)}  P_1 P_2$ decay is divided into two stages. $B_{(s)}$ meson decays to $D_{(s)}^* P_2$ firstly and $D_{(s)}^*$ decays to $D_{(s)}  P_1$ subsequently. The first decay $B_{(s)} \to D_{(s)}^* P_2$ is a weak decay induced by $b \rightarrow \,c\, q\, \bar{u} \, \, (q=d, s)$ at quark level and secondary decay $D_{(s)}^*\to D_{(s)}  P_1$ proceeds via strong interaction.
 In Fig.~\ref{TCE} we list the topological diagrams of these decays under the framework of quasi-two-body decay, including the color-favored tree emission diagram $T$ (a), color-suppressed tree emission diagram $C$ (b) and $W$-exchange diagram $E$ (c), which are specified by topological structures of the weak interaction.
We will not consider processes induced by $b \rightarrow \,u\, q\, \bar{u} \, \, (q=d, s)$
transitions as they are Cabibbo-Kobayashi-Maskawa (CKM)-suppressed and there are
no enough experimental data to do global fit for their nonfactorizable contributions
as done in the $b \to c$ transition case.

The two body charmless $B$ decays have been proved factorization at high precision \cite{Bauer:2000yr}.
Large non-factorizable contribution has been found in the two-body $B$ decays
  with charmed meson final states \cite{ijmpa24-5845,prd69-094018}, which we have illustrated in the
  fourth paragraph of Sec.\ref{Introduction}.
 The matrix elements for the topological diagram $T$, $C$ and $E$  ($C$ and $E$ include large non-factorizable contribution) of $B_{(s)} \to D_{(s)}^* P_2 $, derived by the FAT approach~\cite{Zhou:2015jba}, are given  as
 (see Eqs.(8), (9) and (10) of Ref.~\cite{Zhou:2015jba})
\begin{align}
\begin{aligned}
T^{D^{*}P_2}&=\sqrt2{G_{F}}V_{cb}V_{uq}^{*}a_{1}(\mu)f_{P_2}m_{D^{*}}A_{0}^{B\to D^{*}}(m_{P_2}^{2})(\varepsilon^{*}_{D^{*}}\cdot p_{B}),
\\
C^{D^{*}P_2}&=\sqrt2{G_{F}}V_{cb}V_{uq}^{*} f_{D^{*}}m_{D^{*}}F_{1}^{B\to P_2}(m_{D^{*}}^{2})(\varepsilon^{*}_{D^{*}}\cdot p_{B})\chi^{C}e^{i\phi^{C}},\\
E^{D^{*}P_2}&=\sqrt2{G_{F}}V_{cb}V_{uq}^{*} m_{D^{*}}f_{B}\frac{ f_{D^{*}_{(s)}}f_{P_2}}{f_{D}f_{\pi}}\chi^{E}e^{i\phi^{E}}  (\varepsilon^{*}_{D^{*}}\cdot p_{B})\, .
\end{aligned}
\end{align}
In these equations, for simpler notation the subscript (s) in $D^{*}$ has been kept only in $f_{D_{(s)}^*}$.
This notation is also applied in the following equations involved $D^{*}$, such as
eq.(\ref{Rs}), eq.(\ref{width}) and Eq.(\ref{massDstar}).
$\varepsilon^{*}_{D^{*}}$ is the polarization vector of the $D^*$.
$f_{P_2}$ and $f_{D_{(s)}^*}$ are the decay constants of the corresponding meson $P_2$ and $D_{(s)}^*$.
$ A_0^{B D^*} (m_{P_2}^2) $ and $F_1^{B P_2} (s)$ stand for the vector form factors of $B_{(s)} \to D_{(s)}^*$ and $B_{(s)} \to P_2$ transitions. The form factors are $Q^2$ dependent, and we use the expression in the pole model,
\begin{equation}\label{eq:ffdipole}
F_{i}(Q^{2})={F_{i}(0)\over 1-\alpha_{1}{Q^{2}\over m_{\rm pole}^{2}}+\alpha_{2}{Q^{4}\over m_{\rm pole}^{4}}},
\end{equation}
where $F_{i}$ represents $F_{1}$ and $A_{0}$, and $m_{\rm pole}$ is the mass of the corresponding pole state, such as $B$ for $A_{0}$, and $B^{*}$ for $F_{1}$. 
$ a_1({\mu})$ is the effective Wilson coefficient for the factorizable emission diagram $T$. $\chi^{C(E)}$ and $\phi^{C(E)}$
denote the magnitude and associate phase of $C$($E$) diagram, which are universal and are extracted from the experimental data.

 The intermediate vector $D^*_{(s)}$ resonance is described by the Breit-Wigner formalism\cite{Cheng:2013dua}:
\begin{align}\label{Rs}
R(s)=\frac{1}{s-m_{D^*}^{2}+i m_{D^*} \Gamma_{D^*}(s)}\, ,
\end{align}
where the invariant mass square $s=(p_D+p_1)^2$ with
$p_D$ and $p_1$ denoting the four momenta of the $D$ and $P_1$ mesons, respectively.
The width of $D^*$ meson is energy dependent     \cite{Aaij:2016fma}
\begin{align}\label{width}
\Gamma_{D^*}(s)=\Gamma_{0}\left(\frac{q}{q_{0}}\right)^{3}\left(\frac{m_{D^{*}}}{\sqrt{s}}\right) X^{2}\left(q r_{\mathrm{BW}}\right)\, .
\end{align}
The Blatt-Weisskopf barrier factor is given as~\cite{BW-factor1952}
\begin{equation}
X\left(q r_{\mathrm{BW}}\right)=\sqrt{[1+\left(q_{0} r_{\mathrm{BW}}\right)^{2}]/[1+\left(q r_{\mathrm{BW}}\right)^{2}}]
\end{equation}
with barrier radius $r_{BW} = 4.0~\mathrm{GeV}^{-1}$~\cite{Aaij:2015vea, Aaij:2015kqa, Aaij:2016fma}.
$q=\frac{1}{2} \sqrt{\left[s-\left(m_{D}+m_{P_1}\right)^{2}\right]\left[s-\left(m_{D}-m_{P_1}\right)^{2}\right] / s}$, is the magnitude of the momentum of the final state $D_{(s)} $ or $P_1$ in the rest frame of the $D^*$.
The momentum $q_0$ is the value of $q$ when the invariant mass is equal to the pole mass of the resonance, $s = m^2_{D^*}$.
When a pole mass locates outside the kinematics region, e.g. $m_{D^{*}}<m_{D}+m_{P_1}$,
it will be replaced with an effective mass, $m^{\mathrm{eff}}_{D^*}$, given by the ad hoc formula
\cite{Aaij:2014baa, prd91-092002, Aaij:2016fma},
\begin{align}\label{massDstar}
 m_{D^*}^{\text {eff }}\left(m_{D^*}\right)=m^{\min }+\left(m^{\max }-m^{\min }\right)\left[1+\tanh \left(\frac{m_{D^*}-\frac{m^{\min }+m^{\max }}{2}}{m^{\max }-m^{\min }}\right)\right]\, ,
 \end{align}
 where $m^{\max }$ and $m^{\min }$ are the upper and lower boundaries of the kinematics region, respectively.
The width $\Gamma_{0}$ is the full width of a resonance $D^*$, i.e., the width in Eq.(\ref{width})
when $s = m^2_{D^*}$.
The full width of charged resonance is measured as $\Gamma\left(D^{*+}\right)=83.4 \pm 1.8  ~ \mathrm{keV}$~ \cite{Zyla:2020zbs}.
There is no measurement for neutral $D^*$ meson and we will use the theoretical value obtained by by the light-cone sum rules $\Gamma\left(D^{*0}\right)=55.4\pm 1.4 ~ \mathrm{keV}$ \cite{Li:2020rcg}.
 There is no direct experimental measurement for the full width of $D_s^*$ meson~\cite{Zyla:2020zbs}.
On the theoretical side,   large discrepancies exist, such as
 $\Gamma_{D^{\ast}_s \to D_s \gamma} = 0.066\pm0.026$ keV
in lattice QCD \cite{Donald:2013sra},
$\Gamma_{D^{\ast}_s \to D_s \gamma} = 0.59\pm0.15$ keV in QCD sum rules \cite{Yu:2015xwa}
and $\Gamma_{D^{\ast}_s \to D_s \pi^0} = 8.1_{-2.6}^{+3.0}$ eV in heavy meson chiral perturbation \cite{Yang:2019cat}.
Therefore, we will use the value of 1 keV  to include the largest uncertainty.
To be more clearly, the full widths of these resonance states $D_{(s)}^*$, in addition with
their masses are list in Table \ref{tab:mass and width}.

\begin{table}
\caption{The masses (MeV) and full widths (KeV) of resonance states $D_{(s)}^*$ . }
\vspace{3mm}
\label{tab:mass and width}
\begin{tabular}{ccccccccc|}
\hline
$m_{D^{* \pm}}~~~~~$ & $m_{D^{* 0}}~~~~~$  & $m_{D_s^{* \pm}}~~~~~$ &
$\Gamma_{D^{* \pm}}~~~~~$ &$\Gamma_{D^{* 0}}~~~~~$ & $\Gamma_{D_s^{* \pm}}$&
\\\hline
$2010 ~~~~~$ & $2007~~~~~$ &$2112~~~~~$&$83.4 \pm 1.8 ~~~$&$55.4\pm 1.4~~~$&$1$ &
\\
\hline
\end{tabular}
\end{table}

 The strong decay $\left\langle  P_1 \left(p_{1}\right) D \left(p_{D}\right) | D^{*}(p^*)\right\rangle$
 with the momentum of $D^{*}$ denoted by $p^*$
  is parametrized as a strong coupling constant
  $g_{ D^* D P_1}$, which can be extracted from the partial width
 $\Gamma_{D_{(s)}^*  \to D_{(s)}  P_1}$ by
\begin{align}
\Gamma_{D_{(s)}^*  \to D_{(s)}  P_1}= \frac{p_{P_1}^{3}}{24 \pi m_{D_{(s)}^*}^{2}} g_{D_{(s)}^* D_{(s)} P_1}^{2}\, .
\end{align}
$g_{D^{*} D \pi}$ is determined  precisely as $g_{D^{*} D \pi}=16.92 \pm 0.13 \pm 0.14$~\cite{Lees:2013uxa, Lees:2013zna}
by using the isospin relation $g_{D^{*} D^{0} \pi^{+}}=-\sqrt{2} g_{D^{*} D^{+} \pi^{0}}$ and
the total decay width of the $D^{*+}$.
The other strong couplings can be related with $g_{D^{*} D \pi}$ through the universal coupling $\hat g $
 \cite{Lees:2013uxa, Colangelo:2002dg},
\begin{align}
\hat g\,=\frac{g_{D^{*} D \pi(K)} f_{\pi}}{2 \sqrt{m_{D}^{*} m_{D}}}=\frac{g_{D_{s}^{*} D K} f_{K}}{2 \sqrt{m_{D_{s}}^{*} m_{D}}}\, .
\end{align}
With this universal relation, one gets $g_{D_s^{*} D K}=14.6 \pm 0.06 \pm 0.07$ and $g_{D^{*} D_s K}=14.6 \pm 0.10 \pm 0.13$, which agree very well with the values $g_{D_s^{*} D K}=14.6 \pm 1.7$ and $g_{D^{*} D_s K}=14.7 \pm 1.7$ measured by the CLEO collaboration \cite{Anastassov:2001cw}
and $g_{D_s^{*} D K}=15.2$ and $g_{D^{*} D_s K}=15.2$ calculated by quark model
\cite{Becirevic:1999fr}.

Combing them together, the decay amplitudes of each topological diagrams shown in Fig.\ref{TCE} are given as follows.
\begin{align}\label{Tamp}
\begin{aligned}
&T=\left\langle P_1 \left(p_{1}\right) D \left(p_{D}\right)\left|(\bar{c} b)_{V-A}\right| B (p_B) \right\rangle
\left \langle P_2 \left(p_{2}\right)\left|(\bar{q} u)_{V-A}\right| 0\right\rangle\\
&=\frac{\left\langle  P_1 \left(p_{1}\right) D \left(p_{D}\right) | D^{*}(p^*)\right\rangle}{s-m_{D^*}^{2}+i m_{D^*} \Gamma_{D^*}(s)}
\left\langle D^* (p^*) \left|(\bar{c} b)_{V-A}\right| B(p_B) \right\rangle\left \langle P_2 \left(p_{2}\right)\left|(\bar{q} u)_{V-A}\right| 0\right\rangle\\
&=p_2 \cdot\left(p_{1}-p_{D}\right)\, \sqrt 2 G_{F} V_{c b} V_{u q}^{*} f_{P_2}\, m_{D^*} A_0^{B D^*} (m_{P_2}^2)\, a_1({\mu})\, \frac{g_{ D^* D P_1}} {s-m_{D^*}^{2}+i m_{D^*} \Gamma_{D^*}(s)} \, ,
\end{aligned}
\end{align}

\begin{align}\label{Camp}
\begin{aligned}
&C=\left\langle P_1 \left(p_{1}\right) D \left(p_{D}\right) |(\bar{c} u)_{V-A} |0 \right \rangle
\left\langle P_2 (p_2)|(\bar q b)_{V-A}| B(p_B) \right \rangle \\
&=\frac{\left\langle  P_1 \left(p_{1}\right) D \left(p_{D}\right)| D^{*}(p^*)\right\rangle}{s-m_{D^*}^{2}+i m_{D^*} \Gamma_{D^*}(s)}
\left\langle D^*(p^*) \left|(\bar{c} u)_{V-A}\right| 0 \right\rangle\left \langle P_2 \left(p_{2}\right)\left|(\bar{q} b)_{V-A}\right| B(p_B)\right\rangle\\
&=p_2 \cdot\left(p_{1}-p_{D}\right)\, \sqrt 2 G_{F} V_{c b} V_{u q}^{*} f_{D^*} m_{D^*} F_1^{B P_2} (p^*)\, \chi^{C} \mathrm{e}^{i \phi^{C}}\, \frac{g_{ D^* D P_1}} {s-m_{D^*}^{2}+i m_{D^*} \Gamma_{D^*}(s)} \, ,
\end{aligned}
\end{align}

\begin{align}\label{Eamp}
\begin{aligned}
&E=\left\langle P_1 \left(p_{1}\right) D \left(p_{D}\right) P_2 (p_2) \left|\mathcal{H}_{eff}\right| B(p_B) \right\rangle
\\
&=\frac{\left\langle  P_1 \left(p_{1}\right) D \left(p_{D}\right) | D^{*}(p^*)\right\rangle}{s-m_{D^*}^{2}+i m_{D^*} \Gamma_{D^*}(s)}
\left\langle D^*(p^*) P_2(p_2) \left|\mathcal{H}_{eff} \right| B(p_B)\right\rangle\\
&=p_2 \cdot\left(p_{1}-p_{D}\right)\, \sqrt 2 G_{F} V_{c b} V_{u q}^{*} f_{B} m_{D^*} \frac{f_{D^{*}} f_{P_2}}{f_{\pi} f_{D}}
\chi^{E} \mathrm{e}^{i \phi^{E}} \, \frac{g_{ D^* D P_1}} {s-m_{D^*}^{2}+i m_{D^*} \Gamma_{D^*}(s)} \, .
\end{aligned}
\end{align}
where $p^*=p_1+p_D$.

The total decay amplitude of $B_{(s)}  \to D_{(s)}  P_1 P_2$ can be written as
\begin{align}
\left\langle D_{(s)} \left(p_{D}\right) P_1 \left(p_{1}\right)  P_2 (p_2)  \left| \mathcal{H}_{eff} \right| B_{(s)}(p_B) \right\rangle
= p_2 \cdot\left(p_{1}-p_{D}\right) \mathcal{A}(s)\, ,
\end{align}
where $\mathcal{A}(s)$ represents the summation of amplitudes in Eq.(\ref{Tamp}-\ref{Eamp}) with the factor
$p_2 \cdot\left(p_{1}-p_{D}\right)$ taken out.
The differential width of $B_{(s)}  \to D_{(s)}  P_1 P_2$ is
\begin{align}\label{dwidth}
\begin{aligned}
d \, \Gamma
=&d\, s  \,  \frac{1}{(2 \pi)^3}\,
\frac{(\left|\mathbf{p_1} \| \mathbf{p}_{2}\right|\,)^3}{24 m_B^3}\, | \mathcal{A}(s)|^2\, , \\
\end{aligned}
\end{align}
where $|\mathbf{ p_{1}} |$ and $|\mathbf{ p_{2}} |$ represent the magnitudes of the
momentum $p_1$ and $p_2$, respectively. In the rest frame of the $D_{(s)}^*$ resonance, their expressions are
\begin{align}
|\mathbf{ p_{1}} |=&\frac{1}{2\, \sqrt s}\,\sqrt{ \left[s-(m_D+m_{P_1})^2 \right] \, \left[s-(m_D-m_{P_1})^2 \right]}\, ,\nonumber\\
|\mathbf{ p_{2}} |=&\frac{1}{2\, \sqrt s}\,
\sqrt{ \left[(m_B^2-m_{P_2}^2)^2 \right] \, -2(m_B^2+m_{P_2}^2)^2 \,s +s^2}\, .
\end{align}
where $|\mathbf{ p_{1}} |=q$.
\section{Numerical results and discussion}\label{sec:3}

The input parameters in this work contain
the CKM matrix elements, decay constants, transition form factors, Wilson coefficients and non-pertubative  parameters of $C (E)$ topological diagrams $\chi^{C(E)}$ and $\phi^{C(E)}$.
We use Wolfenstein parametrization of the CKM matrix with the
 Wolfenstein parameters as\cite{Zyla:2020zbs}:
$$ \lambda=0.22650 \pm 0.00048,~~~A=0.790^{+0.017}_{-0.012},~~~\bar \rho=0.141^{+0.016}_{-0.017},~~~
\bar \eta=0.357 \pm 0.01,$$
which lead to $V_{cb} V^*_{ud}=0.0395^{+0.0009}_{-0.0006}$ and  $V_{cb} V^*_{us}=0.0092\pm 0.0002$.
 The decay constants of light pseudoscalar mesons and vector charm mesons $D^*_{(s)}$
 and transition form factors of $B$ meson decays at recoil momentum square
 $Q^2=0$ are listed in Table \ref{tab:decay constants} and Table \ref{tab:ff}, respectively.
The decay constants of $\pi, K,D$ and $B$ are used from the Particle Data Group (PDG) \cite{Zyla:2020zbs}.
There are no experimental data for decay constants of $D_s$, $D^*_{(s)}$ and $B_s$
and all form factors used here. Here we use the same theoretical values
as in the previous work by two of us (S-H Z, C-D L) with other colleagues~\cite{Zhou:2015jba},
with 5\% uncertainty kept for decay constants and 10\% uncertainty for form factors.
Similarly, we will also use the dipole parametrization to describe the $Q^2$ dependence of form factors.
\begin{table}
\caption{The decay constants of light pseudoscalar mesons and vector mesons  (in unit of MeV). }
\vspace{3mm}
\label{tab:decay constants}
\newsavebox{\tablebox}
\begin{lrbox}{\tablebox}
\centering
\begin{tabular}{ccccccccccc|}
\hline
$f_{\pi}$ & $f_{K}$  & $f_{D}$ & $f_{D_s}$ &$f_{D^*}$ & $f_{D_s^*}$ &  $f_{B}$ & $f_{B_s}$ &
\\\hline
$130.2 \pm 1.7$ & $155.6 \pm 0.4$ &$211.9 \pm 1.1$&$258 \pm 12.5 $&$220\pm 11$&$270\pm 14$& $190.9 \pm 4.1$ &
 $225 \pm 11.2$ &
\\
\hline
\end{tabular}
\end{lrbox}
 \scalebox{1}{\usebox{\tablebox}}
\end{table}

\begin{table}
\caption{The transition form factors at $Q^2=0$ and dipole model parameters used in this work. }\label{tab:ff}
\vspace{3mm}
\centering
\begin{tabular}{|c|c|c|c|c|c|c|c|c|c|c|c|c|c|}
\hline&
$~~F_{1}^{B\to\pi}~~$&
$~~F_{1}^{B\to K}~~$&
$~~F_{1}^{B_{s}\to K}~~$&
$~~A_{0}^{B\to D^{*}}~~$&
$~~A_{0}^{B_{s}\to D^{*}_{s}}~~$\\
\hline
$~F(0)~$&
0.28&
0.33&
0.29&
0.56&
0.57\\
$\alpha_1$&
0.52&
0.54&
0.57&
2.44&
2.49\\
$\alpha_2$&
0.45&
0.50&
0.50&
1.98 &
1.74 \\
\hline
\end{tabular}
\end{table}
The Wilson coefficients $C_1$ and $C_2$ at scale $\mu=m_b/2$ are $-0.287$ and $1.132$, respectively.
Then the effective Wilson coefficients $a_1$ is $1.036$.
The non-pertubative contribution parameters
 $\chi^{C(E)}$ and $\phi^{C(E)}$ extracted from experimental data by the fit performed in~\cite{Zhou:2015jba} are
\begin{align}
\begin{aligned}
\chi^{C}&=0.48 \pm 0.01, ~~~~~ \phi^{C}=\left(56.6_{-3.8}^{+3.2}\right)^{\degree}, \\
 \chi^{E}&=0.024_{-0.001}^{+0.002},~~~~~ \phi^{E}=\left(123.9_{-2.2}^{+3.3}\right)^{\degree}.
 \end{aligned}
\end{align}

With all the inputs, the branching fractions of $B^{0,+}\to \bar D_{(s)}^* P_2 \to \bar D_{(s)} P_1 P_2$ and $B_s^0\to \bar D_{(s)}^* P_2 \to \bar D_{(s)} P_1 P_2$ can be obtained by integrating the differential width in Eq.(\ref{dwidth}) over the kinematics region. Our numerical results for $B^{0,+}$ and $B_s^0$ decays are collected in Table \ref{result1} and Table \ref{result2}, respectively. 
In our results the errors are estimated with $5\%$ variations of form factors, $10\%$ variations of decay constants and the uncertainties of the other nonpertubative parameters. One can see that the dominating errors are from the uncertainties of form factors. In the tables we also list the intermediate decays as well as the topological contributions represented by the corresponding symbols.
\begin{table}[tb]   
\begin{center}
\caption{The branching ratios of quasi-two-body decays $B_{u, d} \to D_{(s)}^*P_2 \to D_{(s)} P_1 P_2$.
The decays with on-shell effects are denoted by $\mathcal{B}$, and those without on-shell effects are denoted by $\mathcal{B}_v$. The characters $T$, $C$ and $E$ represent the corresponding topological contributions. The uncertainties are from form factors, decay constants and nonperturbative parameters, respectively. }
\vspace{4mm}
\label{result1}   
\begin{tabular}{ l c c c  c} \hline\hline
\quad\quad\quad Decay modes &  ~Amplitudes~  &  ~~~~~ ${{\mathcal B} ~\text{or} ~{\mathcal B_v}}$ & \quad Results & ~~~~~ Units\;  \\
\hline
$ B^0\to D^{\ast -} \pi^+ \to \bar D^0 \pi^- \pi^+$& T+E &$~~~~~{\mathcal B}$
& $~~~~~~1.77^{+0.39+0.05+0.01}_{-0.35-0.05-0.01}$  &$~~~~~~10^{-3}$\; \\
$\hspace{2.4cm} \to D^- \pi^0 \pi^+$       &        &$~~~~~{\mathcal B}$
& $~~~~~~8.03^{+1.74+0.22+0.07}_{-1.57-0.22-0.04}$  &$~~~~~~10^{-4}$\; \\
$\hspace{2.4cm} \to D_s^-K^0 \pi^+$    &         &$~~~~~{\mathcal B}_v$
& $~~~~~~1.64^{+0.36+0.05+0.01}_{-0.32-0.04-0.01}$  &$~~~~~~10^{-5}$\;  \\
$B^0\to D^{\ast -}K^+ \to \bar D^0 \pi^- K^+$  & T      &$~~~~~{\mathcal B}$
& $~~~~~~1.47^{+0.31+0.08+0}_{-0.28-0.08-0}$   &$\,~~~~~~10^{-4}$ \, \\
$\hspace{2.5cm} \to D^- \pi^0 K^+$     &    &$~~~~~{\mathcal B}$
& $~~~~~~6.63^{+1.39+0.03+0}_{-1.26-0.03-0}$   &$\,~~~~~~10^{-5}$ \,\\
$\hspace{2.5cm} \to D_s^-K^0 K^+$     &    &$~~~~~{\mathcal B}_v$
& $~~~~~~1.28^{+0.27+0.01+0}_{-0.24-0.01-0}$   &$\,~~~~~~10^{-6}$ \,\\
\hline
$ B^0\to \bar D^{\ast 0} \pi^0 \to \bar D^0 \pi^0 \pi^0$   &$\frac{1}{\sqrt 2} (E-C)$    &$~~~~~{\mathcal B}$
& $~~~~~~1.39^{+0.30+0.14+0.07}_{-0.27-0.14-0.07}$  &$~~~~~~10^{-4}$\;\\
$\hspace{2.2cm} \to D^- \pi^+ \pi^0$  &       &$~~~~~{\mathcal B}_v$
& $~~~~~~1.07^{+0.23+0.10+0.05}_{-0.21-0.10-0.05}$  &$~~~~~~10^{-5}$\;\\
$\hspace{2.2cm} \to D_s^- K^+ \pi^0$    &     &$~~~~~{\mathcal B}_v$
& $~~~~~~1.60^{+0.35+0.16+0.08}_{-0.31-0.16-0.08}$  &$~~~~~~10^{-6}$\;\\
$B^0\to \bar D^{\ast 0} K^0 \to \bar D^0 \pi^0 K^0$  & C      &$~~~~~{\mathcal B}$
& $~~~~~~2.23^{+0.47+0.23+0.09}_{-0.42-0.22-0.09}$  &$~~~~~~10^{-5}$\; \\
$\hspace{2.3cm} \, \to D^- \pi^+ K^0$     &    &$~~~~~{\mathcal B}_v$
& $~~~~~~1.67^{+0.35+0.17+0.07}_{-0.32-0.16-0.07}$  &$~~~~~~10^{-6}$\; \\
$\hspace{2.3cm} \, \to D_s^- K^+ K^0$     &    &$~~~~~{\mathcal B}_v$
& $~~~~~~2.38^{+0.50+0.24+0.10}_{-0.45-0.23-0.10}$  &$~~~~~~10^{-7}$\; \\
\hline
$ B^+\to \bar D^{\ast 0} \pi^+\; \to \bar D^0 \pi^0 \pi^+$    &T+C     &$~~~~~{\mathcal B}$
& $~~~~~~3.09^{+0.51+0.01+0.08}_{-0.47-0.01-0.09}$  &$~~~~~~10^{-3}$\;\\
$\hspace{2.6cm} \to D^- \pi^+ \pi^+$    &     &$~~~~~{\mathcal B}_v$
& $~~~~~~2.27^{+0.37+0.07+0.06}_{-0.34-0.07-0.06}$  &$~~~~~~10^{-4}$\; \\
$\hspace{2.7cm} \to D_s^- K^+ \pi^+$    &     &$~~~~~{\mathcal B}_v$
& $~~~~~~3.21^{+0.52+0.11+0.09}_{-0.48-0.10-0.10}$  &$~~~~~~10^{-5}$\; \\
$ B^+\to \bar D^{\ast 0} K^+ \to \bar D^0 \pi^0 K^+$    &T+C     &$~~~~~{\mathcal B}$
& $~~~~~~2.35^{+0.39+0.06+0.06}_{-0.36-0.06-0.07}$  &$~~~~~~10^{-4}$\;\\
$\hspace{2.6cm} \to D^- \pi^+ K^+$   &      &$~~~~~{\mathcal B}_v$
& $~~~~~~1.69^{+0.28+0.04+0.04}_{-0.26-0.04-0.05}$  &$~~~~~~10^{-5}$\; \\
$\hspace{2.7cm} \to D_s^- K^+ K^+$    &     &$~~~~~{\mathcal B}_v$
& $~~~~~~2.29^{+0.37+0.06+0.06}_{-0.35-0.06-0.07}$  &$~~~~~~10^{-6}$\;  \\
\hline\hline
\end{tabular}
\end{center}
\vspace{-4mm}
\end{table}

\begin{table}[tb]   
\begin{center}
\caption{The same as table \ref{result1}, but for the quasi-two-body $B_s^0\to D_{(s)}^\ast P_2 \to D P_1 P_2$ decays}
\label{result2}   
\vspace{4mm}
\begin{tabular}{ l c c c c }
\hline\hline
\quad\quad\quad  Decay modes  &  ~Amplitudes~      & ~\; ${{\mathcal B}~\text{or} ~{\mathcal B_v}}$  & \quad Results & ~~~~~ Units\;  \\
\hline
$ B_s^0\to D_s^{\ast -} \pi^+ \to \bar D^0 K^- \pi^+$    & T   &$~~~~~{\mathcal B}_v$
& $~~~~~~4.21^{+0.88+0.11+0}_{-0.80-0.11-0}$   &$~~~~~~10^{-5}$\; \\
$\hspace{2.4cm} \to  D^- \bar K^0 \pi^+$    &     &$~~~~~{\mathcal B}_v$
& $~~~~~~4.07^{+0.85+0.11+0}_{-0.77-0.11-0}$\   &$~~~~~~10^{-5}$\; \\
$ B_s^0\to D_s^{\ast -} K^+ \to \bar D^0 K^- K^+$   &T+E      &$~~~~~{\mathcal B}_v$
& $~~~~~~2.84^{+0.62+0.02+0.03}_{-0.56-0.02-0.02}$    &$~~~~~~10^{-6}$\;\\
$\hspace{2.5cm} \to  D^- \bar K^0 K^+$     &    &$~~~~~{\mathcal B}_v$
& $~~~~~~2.74^{+0.60+0.02+0.03}_{-0.54-0.02-0.02}$    &$~~~~~~10^{-6}$\; \\
\hline
$ B_s^0\to D^{\ast -} \pi^+ \to \bar D^0 \pi^- \pi^+$  &E      &$~~~~~{\mathcal B}$
& $~~~~~~5.98^{+0+0.87+1.04}_{-0-0.83-0.49}$    &$~~~~~~10^{-7}$\;\\
$\hspace{2.4cm} \to  D^- \pi^0 \pi^+$   &      &$~~~~~{\mathcal B}$
& $~~~~~~2.71^{+0+0.39+0.47}_{-0-0.37-0.22}$    &$~~~~~~10^{-7}$\; \\
$\hspace{2.4cm} \to  D_s^-K^0 \pi^+$    &     &$~~~~~{\mathcal B}_v$
& $~~~~~~5.80^{+0+0.84+1.01}_{-0-0.80-0.47}$    &$~~~~~~10^{-9}$\; \\
\hline
$ B_s^0\to \bar D^{\ast 0} \pi^0 \to \bar D^0 \pi^0 \pi^0$ &$\frac{1}{\sqrt 2}  E$ &$~~~~~{\mathcal B}$
& $~~~~~~2.76^{+0+0.30+0.36}_{-0-0.28-0.17}$    &$~~~~~~10^{-7}$\;\\
$\hspace{2.3cm} \to  D^- \pi^+ \pi^0$   &      &$~~~~~{\mathcal B}_v$
& $~~~~~~2.03^{+0+0.40+0.48}_{-0-0.38-0.23}$    &$~~~~~~10^{-8}$\;\\
$\hspace{2.3cm} \to  D_s^- K^+ \pi^0$   &      &$~~~~~{\mathcal B}_v$
& $~~~~~~2.91^{+0+0.42+0.50}_{-0-0.40-0.24}$    &$~~~~~~10^{-9}$\;\\
$ B_s^0\to \bar D^{*0} \bar K^0 \to \bar D^0 \pi^0 \bar K^0$ &C &$~~~~~{\mathcal B}$
& $~~~~~~3.40^{+0.71+0.35+0.14}_{-0.65-0.33-0.14} $   &$~~~~~10^{-4}$ \\
$\hspace{2.3cm} \to  D^- \pi^+ \bar K^0$    &     &$~~~~~{\mathcal B}_v$
& $~~~~~~2.61^{+0.55+0.27+0.11}_{-0.50-0.26-0.11}$    &$~~~~~10^{-5}$ \\
$\hspace{2.3cm} \to  D_s^- K^+ \bar K^0$  &       &$~~~~~{\mathcal B}_v$
& $~~~~~~3.88^{+0.81+0.40+0.16}_{-0.74-0.38-0.16}$    &$~~~~~10^{-6}$\\
\hline\hline
\end{tabular}
\end{center}
\end{table}

 \subsection{The hierarchy of branching fraction}

The hierarchies of branching fractions can be seen clearly in the last columns of Tables~\ref{result1} and \ref{result2}.
Firstly, let us pay attention to those decays  in Table \ref{result1} whose intermediate states can be changed to each other by swapping the ``bachelor" light mesons pion with kaon,
e.g., $ B^0\to D^{\ast -} \pi^+ \to \bar D^0 \pi^- \pi^+$ and $B^0\to D^{\ast -}K^+ \to \bar D^0 \pi^- K^+$.
One can find that the branching ratios of the modes with a pion ``bachelor" meson are one power larger than their corresponding ones with a kaon ``bachelor" meson. The reason is that the pion  ``bachelor" modes are the Cabibbo favored ($V_{ud}$) processes,
while the kaon ``bachelor" modes are Cabibbo suppressed  ($V_{us}$) ones.
 The branching ratios of the Cabibbo favored decays
$ B_s^0\to D_s^{\ast -} \pi^+ \to \bar D^0 K^- \pi^+$  and
$ B_s^0\to \bar D^{*0} \bar K^0 \to \bar D^0 \pi^0 \bar K^0$ shown in Table \ref{result2} are larger than those of the remaining Cabibbo suppressed decays.

Similar to the    dynamics in two body hadronic $B$ decays, the color favored tree ($T$) topological diagram is absolutely dominating. Branching fractions of decays with $T$ diagram are larger than those with only   $C$ or $E$ diagrams. Our results for these branching fractions with $T$ diagram are in good agreement with the PQCD predictions~\cite{Chai:2021kie}. We have   $\left|C \right| > \left|E\right|$ in the FAT approach \cite{Zhou:2015jba} while
$\left|C \right| \sim\left|E\right|$ in the PQCD approach \cite{Li:2008ts}. The $\left|C \right| $ contributes larger in the FAT approach than in the PQCD approach. Therefore, our branching fractions of $B^0\to \bar D^{\ast 0} K^0 \to \bar D^0 \pi^0 K^0$ and  $ B_s^0\to \bar D^{*0} \bar K^0 \to \bar D^0 \pi^0 \bar K^0$ decays with only $C$ diagram are   about two times larger than that in the PQCD approach.

For those decays with exactly the same intermediate state $D^*_{(s)} P_2$,
their hierarchies of branching ratios are completely caused by the strong decays of $D_{(s)}^{*} \to D_{(s)} P_1$. There are two hierarchy sources. One is the BWT effect, the other is the strong couplings. Some of the decays, labeled by $\mathcal{B}$ in the tables,
can proceed by the pole mass dynamics, i.e., the pole mass is larger than
the invariant mass threshold of two final states. The others labeled by $\mathcal{B}_v$ can only happen by the BWT effect.
The branching ratios of ${\cal B}$ mode decays are apparently one to two orders larger than those of $\mathcal{B}_v$ mode decays.
The difference due to strong couplings are also obvious. For instance, the isospin relation leads to $g_{D^{*-} \bar D^0 \pi^{-}}=-\sqrt 2\, g_{D^{*-} D^- \pi^{0}}$.  As a result, the branching ratio of $ B_{(s)}^0 \to \bar D^0 \pi^- \pi^+ (K^+)$ is approximatively two times larger than that of
$ B_{(s)}^0 \to D^- \pi^0 \pi^+ (K^+)$. The strong decays with a pair of $s \bar s$ quarks from the sea   proceed  only by the BWT effect because of the heavy strange quark mass.
Consequently, the corresponding branching fractions are smaller than the others.

\subsection{Dependence of branching fractions on the invariant mass of intermediate state}

\begin{figure}
\begin{center}
\vspace{-1cm}
\scalebox{0.5}{\epsfig{file=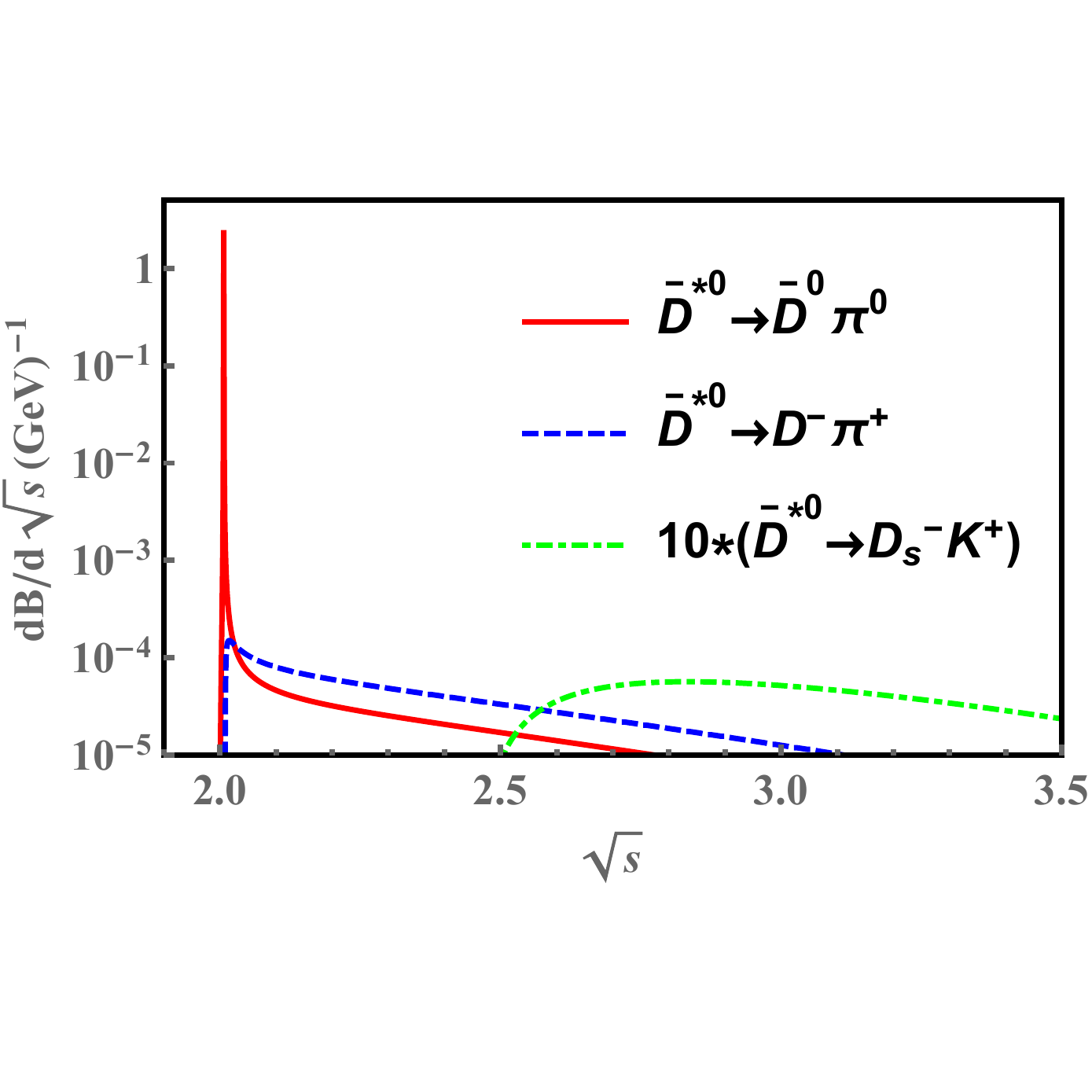}}\quad\quad
\scalebox{0.5}{\epsfig{file=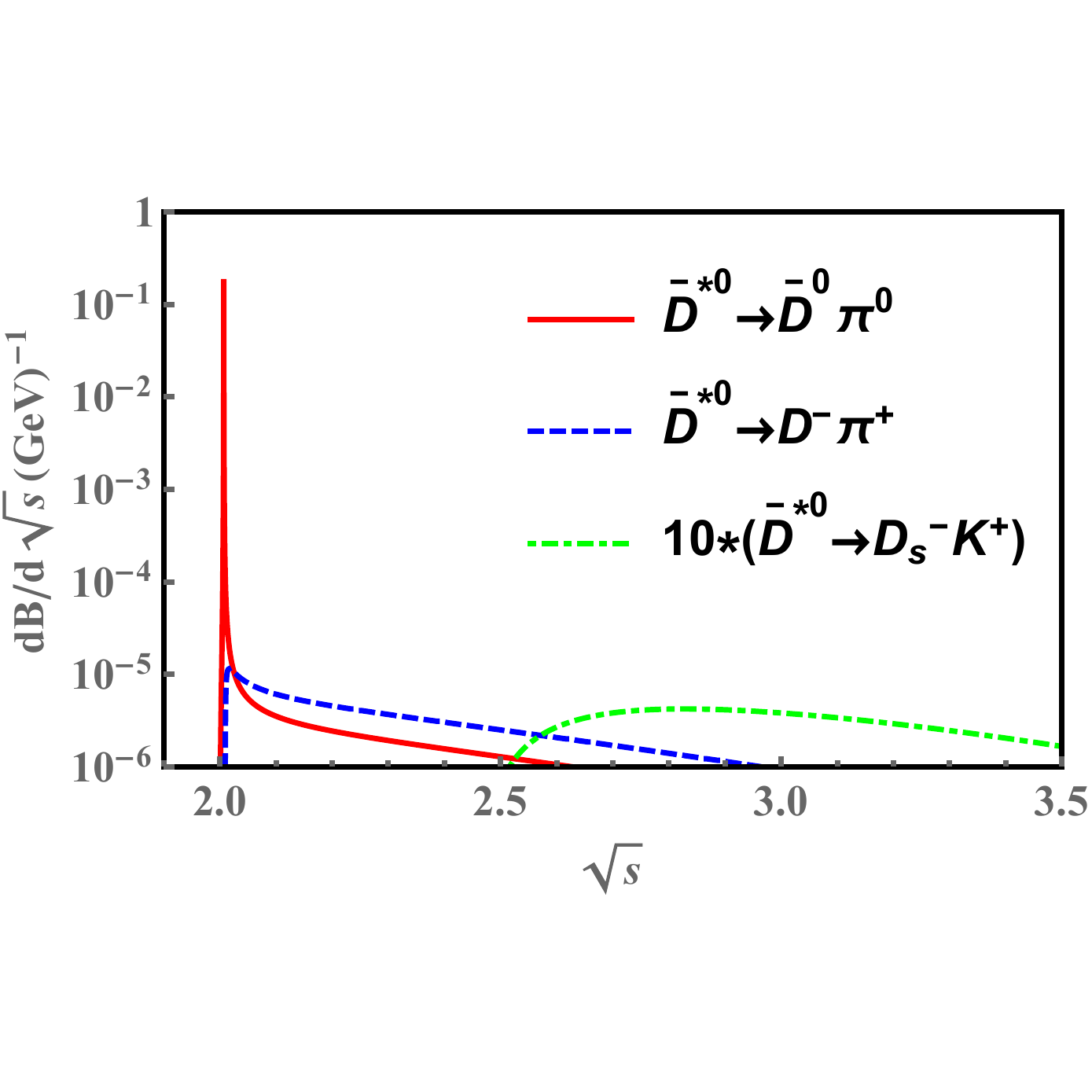}}
\vspace{-15mm}
\caption{The differential branching fractions of
$ B^+\to \bar D^{\ast 0} \pi^+\; \to \bar D^0 \pi^0\,  (D^- \pi^+\, ,  D_s^- K^+)\,  \pi^+$ decays (left pannel)
and corresponding decays with ``bachelor" pion replaced with kaon  (right panel). }
\label{BRs}
\end{center}
\end{figure}

We take the decays $ B^+\to \bar D^{\ast 0} \pi^+\; \to \bar D^0 \pi^0\,  (D^- \pi^+\, ,  D_s^- K^+)\,  \pi^+$
as an example to show the dependence of branching fractions on the invariant mass of $\bar D^0 \pi^0$ pair  (or $D^- \pi^+$, $ D_s^- K^+$), which is depicted in the left diagram of  Fig.\ref{BRs}. To demonstrate the flavor SU(3) symmetry breaking effect, curves of their corresponding decays with the ``bachelor" pion replaced with kaon are plotted in the right diagram of Fig.\ref{BRs}.
The decay with a pole mass dynamical strong decay
$\bar D^{\ast 0} \to \bar D^0 \pi^0 $ and those with the BWT effect strong decays
$\bar D^{\ast 0} \to D^- \pi^+\ $ and $\bar D^{\ast 0} \to D_s^- K^+ $
are represented by solid (red) lines, dash (blue) lines and dot-dashed (green) lines, respectively.
 Data of the decays with $\bar D^{\ast 0} \to D_s^- K^+\, $, which are one order magnitude smaller than those with $\bar D^{\ast 0} \to D^- \pi^+\, $, are multiplied by ten in oder to be shown clearly on the same figure.
 One may notice that the two conspicuous flagpoles located at
 the mass of $\bar D^{\ast 0}$ in Fig.\ref{BRs}. It is the distinguishing characteristic of pole dynamics.
 %
%
Our curves of $ B^+\to \bar D^{\ast 0} \pi^+\; \to \bar D^0 \pi^0\,  (D^- \pi^+\, ,  D_s^- K^+)\,  \pi^+$ (the red solid lines) have higher peaks than that
from the ones in the PQCD approach \cite{Chai:2021kie}, which indicates the pole mass dynamics plays a more important role in our results.

Only a few charmed three-body B decays are measured by the experiments. We collected these channels  in Table \ref{BWT} as well as the results of the PQCD approach.  Our results are consistent with the PQCD ones except the decay $B_s^0 \to D_s^{\ast -} \pi^+ \to \bar D^0 K^- \pi^+$. However, the experimental data of this decay has a very large uncertainty. It may be a good choice to wait for more accurate experimental measurement.
Our calculation of the intermediate decays $B \to D_{(s)}^* P$ is more precise than the PQCD approach,
especially for the color suppressed (C) or exchange decay amplitudes (E) as
this two non-factorizable amplitudes, $C$ and $E$, are fitted from precise experiment data in FAT
not calculated perturbatively. Therefore our theoretical uncertainty is much smaller than the previous PQCD approach.
We also study  the BWT effects in these decays. Although the $B_s$ decay in  Table \ref{BWT} has only BWT contribution, the four decays of $B^{+,0}$ are dominated by pole dynamics. In order to make an effective comparison, the BWT effects of $B^{+,0}$ decays are calculated with a cut of $\sqrt s \geq 2.1 \mathrm{GeV}$, which is also adopted by experiments and the PQCD approach.
Comparing the data in Tables~\ref{BWT} and \ref{result1},
one can see that the BWT effect contributes  about $5\%$ in $B^0$ meson decays.
For example, the centre value of the branching ratio of $B^0\to D^{\ast -} \pi^+ \to \bar D^0 \pi^- \pi^+$
 in Tables~\ref{BWT}  by a cut of $\sqrt s \geq 2.1 \mathrm{GeV}$ in FAT is $10.26\times 10^{-5}$
and the one in the whole physical region of the $ \bar D^0 \pi^- $
 invariant mass \ref{result1} is $1.77\times 10^{-3}$, which lead to the percentage of BWT effect is
 $10.26\times 10^{-5}/1.77\times 10^{-3} \sim 5\%$.

\begin{table}[t]
\caption{The comparison of BWT effects between FAT, PQCD predictions and available experimental measurements,
where ${\mathcal B}_{\mathrm{FAT, PQCD}}^{{\rm cut}}$ represents a cut on the invariant mass for $B^{+, 0}$ decays.}
 \vspace{2mm}
   \begin{lrbox}{\tablebox}
   \centering
   \begin{tabular}{c c c l }
			\hline\hline
			Decay modes
			  & \quad\quad${\mathcal B}_{\rm {FAT}}^{({\rm cut})} (10^{-5})$ \quad\quad	
			 &\quad\quad${\mathcal B}_{\rm {PQCD}}^{({\rm cut})} (10^{-5})$ \cite{plb791-342} \quad\quad
			 &\quad\quad Data $(10^{-5})$ \quad\quad \\
			\hline
			$B^+\to \bar D^{\ast 0}\pi^+ \to D^-\pi^+\pi^+$ &
			$18.8^{+3.16}_{-2.96}$ &
			$19.2^{+8.80}_{-6.20} $  &
			 $22.3\pm3.20 $ \,\cite{prd69-112002}  \\
			 ~& ~ & ~ & $10.9 \pm 1.8$\, ~\cite{prd79-112004} \\
			 ~& ~ & ~ & $10.9 \pm 2.7$ \, ~\cite{Aaij:2016fma} \\
			$B^+\to \bar D^{\ast 0}K^+\to D^-\pi^+K^+$
			& $1.39^{+0.24}_{-0.21} $
			 & $1.48^{+0.68}_{-0.47} $
			&$0.56 \pm 0.23 $ \, \cite{prd91-092002} \\
			$ B^0\to D^{\ast -} \pi^+ \to \bar D^0 \pi^- \pi^+$&
			$10.26^{+2.25}_{-2.02}$&$8.7^{+4.5}_{-2.9}$
			& $8.8 \pm 1.3$ \, \, \, \cite{Kuzmin:2006mw} \\
			 ~& ~ & ~ &$ 7.8 $~~~~~~~~~~~\, \cite{Aaij:2015sqa} \\
				$B^0\to D^{\ast -}K^+ \to \bar D^0 \pi^- K^+$&
			$0.83^{+0.17}_{-0.15}$&$0.72^{+0.36}_{-0.24}$
			& $0.81 \pm 0.38$ \, \cite{Aaij:2015kqa}\\
			$B_s^0 \to D_s^{\ast -} \pi^+ \to \bar D^0 K^- \pi^+$
			& $4.21^{+0.89}_{-0.81}$&$1.90^{+1.01}_{-0.68}$
			& $4.70 \pm 4.38$ \, \cite{prd90-072003} \\
			\hline\hline
		\end{tabular}
\end{lrbox}
\label{BWT}
 \scalebox{1}{\usebox{\tablebox}}
\end{table}

Comparing the BWT effect between
$ B^+\to \bar D^{\ast 0} \pi^+\; \to D^- \pi^+\, \pi^+$  and
$ B^+\to \bar D^{\ast 0} \pi^+\; \to D_s^- K^+\, \pi^+$,
 we find that the peaks of the green (dot-dashed) lines locate farther from the pole mass of $\bar D^{\ast 0}$ than the peaks of the blue (dashed) lines in Fig.\ref{BRs}. The reason is that $D_s^- K^+$ has much larger threshold mass than $D^- \pi^+$.
 It can also be seen that the green (dot-dashed) line is smoother than the blue (dashed) line,
 which is very steep in the vicinity of the pole mass. Our calculation shows that the BWT effect in $ B^+\to \bar D^{\ast 0} \pi^+\; \to D^- \pi^+\, \pi^+$ is about $4\%$.

 To check the dependence of the BWT effect on the widths of the intermediate resonance, we can compare those decays with different intermediate states. For instance, we find that contribution of the BWT effect is $(1.64^{+0.36}_{-0.32})\times 10^{-5}$ for
  $ B^0\to D^{\ast -} \pi^+\, \to D_s^-K^0 \pi^+$, $(3.21^{+0.54}_{-0.50}) \times 10^{-5}$ for $B^+ \to \bar{D}^{\ast 0} \pi^+ \to D_s^- K^+ \pi^+$, and $(4.21^{+0.89}_{-0.81})\times 10^{-5}$ for $B_s^0 \to D_s^{\ast -} \pi^+ \to \bar D^0 K^- \pi^+$. Their BWT effects are at the same order
 even though the widths of the resonances vary widely with $\Gamma_{D^{*+}}=83.4 \pm 1.8  ~ \mathrm{keV}$,
 $\Gamma_{D^{*0}}=55.3\pm 1.4 ~ \mathrm{keV}$ and $\Gamma_{D^{\ast -}_s} \simeq 1$ keV.
It indicates that the BWT effects in these decays are not very sensitive to
the widths of resonances, which is confirmed by the PQCD approach \cite{plb791-342, Chai:2021kie}. This can be explained by the behaviour of the Breit-Wigner propagators in the kinematics regions of these decays,
where the real parts of their denominators are much larger than the imaginary parts.
Taking $D_s^{\ast -}\to \bar D^0 K^-$ as an example, the kinematics region starts from $\sqrt{s}=2.359\, \mathrm{GeV}$, while $\Gamma_{D^{\ast -}_s} \simeq 1$keV. It is obvious that $\mid m_{D_{s}^{*-}}^{2}-s\mid \gg \mid m_{D_{s}^{*-}} \Gamma_{D_{s}^{*-}}(s)\mid$.

  \subsection{Flavor $SU(3)$ symmetry breaking}

Now, we turn to discuss the flavor-SU(3) symmetry breaking effect in the quasi-two-body decays $B^+ \to \bar D^{*0}\pi^+ (K^+) \to \bar D^0\pi^0\pi^+ (K^+)$ and $B^0 \to D^{*-}\pi^+ (K^+) \to \bar D^0\pi^- \pi^+ (K^+)$ with different bachelor particles $\pi$ and K. The ratio of the branching fractions,
 \begin{align}
 R_{D^{*-}}&=\frac{\mathcal{B}\left(\bar{B}^{0} \rightarrow D^{*-} K^{+} \rightarrow \bar D^{0} \pi^{-} K^{+}\right)}{\mathcal{B}\left(\bar{B}^{0} \rightarrow D^{*-} \pi^{+} \rightarrow \bar D^{0} \pi^{-} \pi^{+}\right)}\, , \quad
 R_{D^{* 0}}=\frac{\mathcal{B}\left( B^+\to \bar D^{\ast 0} K^+\; \to \bar D^0 \pi^0  K^+\right)}
 {\mathcal{B}\left( B^+\to \bar D^{\ast 0} \pi^+ \; \to \bar D^0 \pi^0 \pi^+\right)}\, ,
 \end{align}
can be used to test SU(3) symmetry breaking. Substituting the results in Table \ref{result1}  into the above equations, one gets
 \begin{align}\label{SU3R}
 R_{D^{*-}}&=0.0831^{+0.0008}_{-0.0006}\, , \quad
 R_{D^{* 0}}=0.0761\pm 0.0001\, .
 \end{align}
$ R_{D^{*-}}$ is consistent with the value $\left(0.081_{-0.002}^{+0.000}\left(\omega_{B}\right)_{-0.000}^{+0.001}\left(a_{D \pi}\right)\right)$ in the PQCD approach \cite{plb791-342}, $(7.76 \pm 0.34 \pm 0.29) \%$
measured by the BABAR collaboration \cite{Aubert:2005yt} and $(0.074 \pm 0.015 \pm 0.006)$ by the Belle collaboration \cite{Abe:2001waa}.
$R_{D^{* 0}}$ also agrees well with the PQCD's result of $R_{D^{* 0}}=0.077_{-0.001}^{+0.000}\left(\omega_{B}\right)_{-0.001}^{+0.000}\left(\omega_{D \pi}\right)$ \cite{plb791-342} and the experimental data $R_{D^{* 0}}=0.0813 \pm 0.0040 (\text {stat })_{-0.0031}^{+0.0042}( \text {syst} )$ \cite{Aubert:2004hu}.
The above ratios $R_{D^{*-}} \simeq R_{D^{* 0}} \simeq |V_{us} /V_{ud} |^2 \times (f_K/ f_\pi)^2 = 0.076$, which indicates that the
source of SU(3) asymmetries are
mainly from decay constants or weak transition form factors.

It is expected that the $SU(3)$ symmetry breaking effect in the quasi-two-body decays should be equal to the breaking effect calculated with
their intermediate two-body decays
\begin{align}\label{ratio}
  R_{D^{* 0}}=\frac{\mathcal{B}\left( B^+\to \bar D^{\ast 0} K^+\; \to \bar D^0 \pi^0  K^+\right)}
 {\mathcal{B}\left( B^+\to \bar D^{\ast 0} \pi^+ \; \to \bar D^0 \pi^0 \pi^+\right)}\,
\approx \frac{\mathcal{B}\left(B^{+} \rightarrow \bar D^{* 0} K^{+} \right)}{\mathcal{B}\left(B^{+} \rightarrow \bar D^{* 0} \pi^{+}\right)}\, .
 \end{align}
This conclusion can be obtained with the narrow width approximation, under which one has
\begin{align}
\mathcal{B}\left[B^+ \to \bar D^{* 0} \pi^+(K^+) \to \bar D^{0} \pi^0 \pi^+(K^+)\right] \approx \mathcal{B} \left[B^+ \to \bar D^{* 0} \pi^+(K^+)\right] \cdot \mathcal{B}\left[\bar D^{* 0} \to \bar D^{0} \pi^0 \right].
 \end{align}

 Eq.~(\ref{ratio}) can be checked numerically. Using the branching fractions calculated with the FAT approach~\cite{Zhou:2015jba}
 \begin{align}
 \mathcal{B}\left(B^{+} \rightarrow \bar D^{* 0} K^{+} \right)&=(3.8 ^{+0.6}_{-0.4})\times 10^{-4} \, , \nonumber\\
 \mathcal{B}\left(B^{+} \rightarrow \bar D^{* 0} \pi^{+} \right)&=(50.7 ^{+8.1}_{-8.2})\times 10^{-4} \, ,
  \end{align}
one gets
  \begin{align}
 \frac{\mathcal{B}\left(B^{+} \rightarrow \bar D^{* 0} K^{+} \right)}{\mathcal{B}\left(B^{+} \rightarrow \bar D^{* 0} \pi^{+}\right)}=0.0750\pm 0.0001\, .
  \end{align}
This value agrees with the $R_{D^{* 0}}$ in Eq.(\ref{SU3R}), which indicates that the narrow width approximation works very well in these decays.

\begin{figure}
\begin{center}
\vspace{-1cm}
\scalebox{0.5}{\epsfig{file=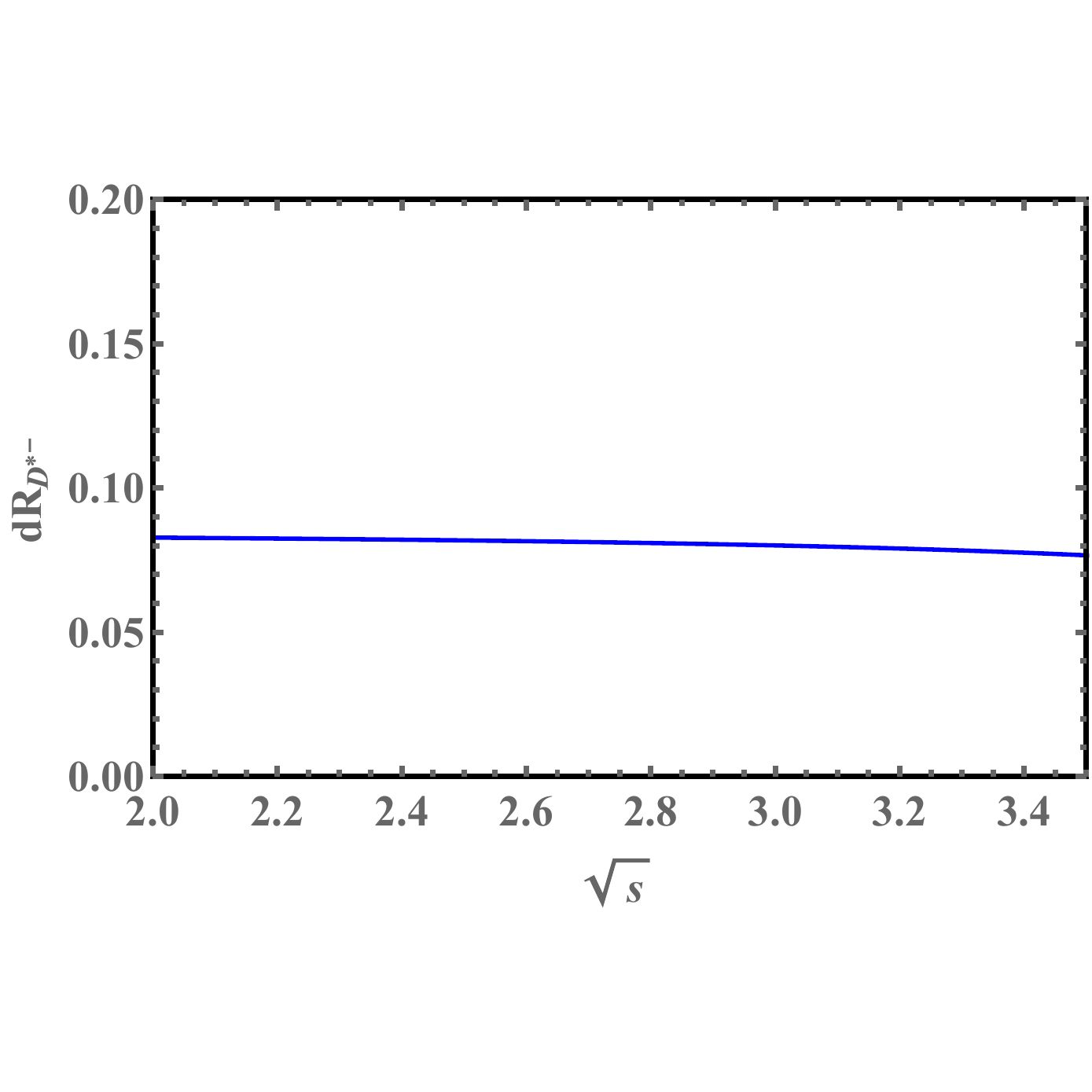}}\quad
\scalebox{0.5}{\epsfig{file=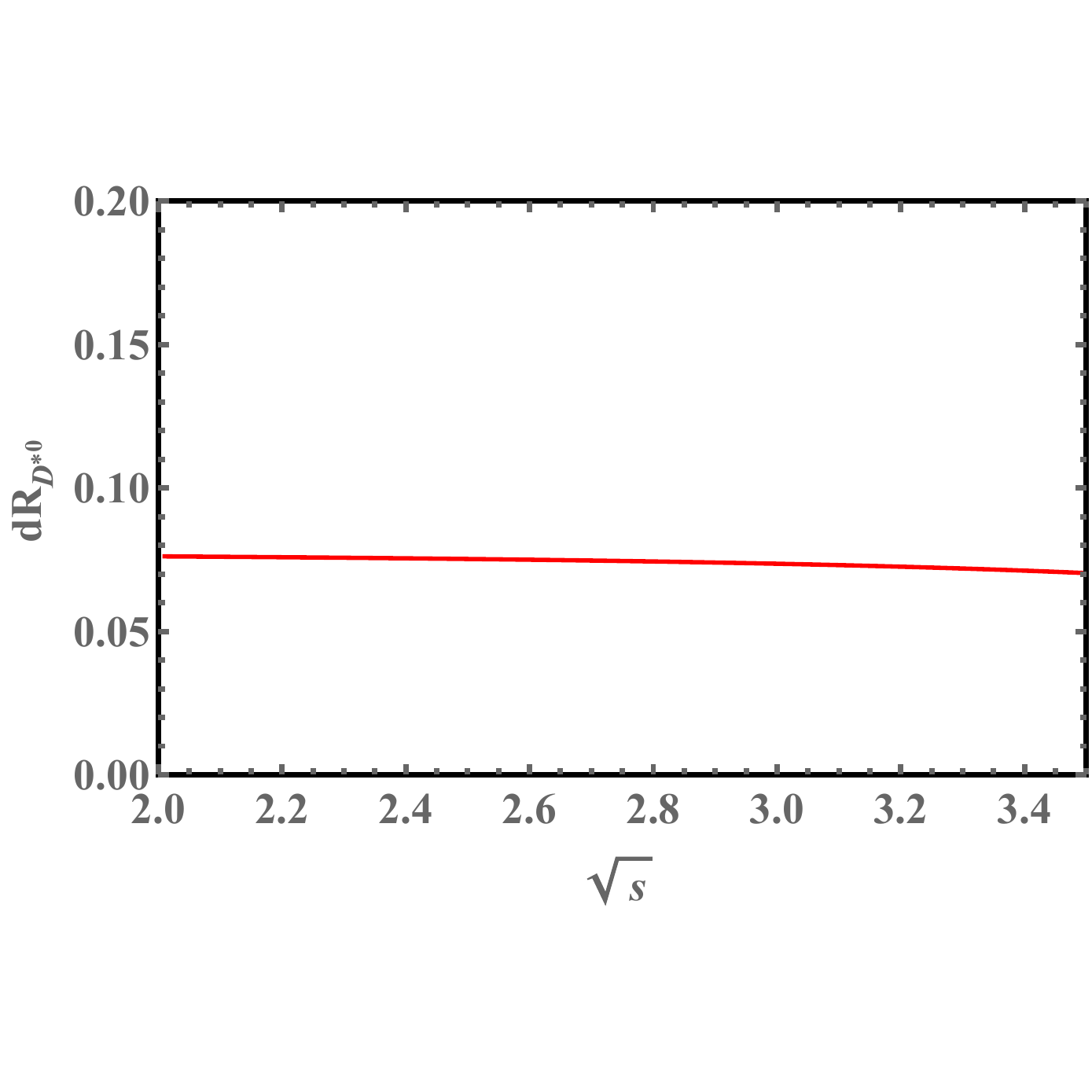}}
\vspace{-15mm}
\caption{The plotted curves are the ratios of corresponding differential branching fractions as functions of
the invariant mass of $\bar D^0\pi^-$ or $\bar D^0\pi^0$. These ratios could be defined as
$\mathrm{d} \, R_{D^{*-}} $ or $\mathrm{d} \, R_{D^{*0}} $ which could be the y labels.}
\label{RDstar}
\end{center}
\end{figure}

Besides, the above conclusion can also be checked by  the local $SU(3)$ breaking effects, which are defined as the ratios of corresponding differential branching fractions  on the invariant mass of $\bar D^0\pi^-$ or $\bar D^0\pi^0$. Our results are plotted in Fig.\ref{RDstar}. It can be seen that the magnitudes of the $SU(3)$ symmetry breaking at every physical point are approximately equal.

\section{Conclusion}\label{sec:4}

The three-body charmed $B$ meson decays are   dominated
by intermediate resonances, i.e., they proceed via quasi-two-body decays as
$B_{(s)} \to D_{(s)}^* P_2  \to D_{(s)}  P_1 P_2$. The first step of two-body decay is induced by flavor changing weak decays
$b \rightarrow \,c\, q\, \bar{u} \quad(q=d, s)$ and the intermediate resonant state $D_{(s)}^* $
successionally decays into $D_{(s)}  P_1$ via strong interaction.
We utilize the  decay amplitudes  extracted from the two-body charmed $B$ decays to quasi-two-body decays with the intermediate resonance described by the Breit-Wigner propagator. Including all the P-wave resonant states $\bar D_{(s)}^*$, we systematically study these decays. The $B_{(s)} \to D_{(s)}^* P_2  \to D_{(s)}  P_1 P_2$ decays can be divided into two groups: $\mathcal{B}$ and $\mathcal{B}_v$. $\mathcal{B}$ represents the group of decays whose pole mass of the resonance $\bar D_{(s)}^*$ is larger than the threshold mass of producing $\bar D_{(s)}  P_1$. This group of decays are dominated by the pole dynamics, and the main contributions to their branching fractions are from vicinity of the resonances' pole masses.  The others labeled by $\mathcal{B}_v$ can only happen by the BWT effect.
 The BWT effects in $B^0$ decays  are approximately $5\%$. We also found that the Breit-Wigner-tail effects are not sensitive to the widths of their corresponding resonances.

We compare our results of the branching fractions   with the PQCD approach's predictions as well as the experimental data. The largest theoretical uncertainty in the calculation is from the intermediate two-body charmed $B$ meson decays. Since this calculation is done by a global fit to all experimental data, our results of three body $B$ decays have significantly less theoretical uncertainty than the PQCD calculation.  We use the same cut on the kinematics region of those decays proceeding via pole dynamics as   the experiments. Our results agree well with the experimental data. For those ones   without any experimental data, we would like to confront them with the future   more accurate experiment measurements.

The flavor-$SU(3)$ symmetry breaking effect in these decays is studied. Our results $R_{D^{*-}}=0.0831^{+0.0008}_{-0.0006}$ and $R_{D^{* 0}}=0.0761\pm 0.0001$ are perfectly consistent with their experimental values. Besides, the local $SU(3)$-breaking effects, which are defined as ratios of corresponding differential branching fractions, are investigated. Their magnitudes are found not sensitive to the invariant mass of the strong decay final states, which confirms the that the SU(3) asymmetry in three-body decays is dominated by the two-body weak decays.

\section*{Acknowledgments}
We are grateful to Wen-Fei Wang for useful discussion.
The work is supported by the National Natural Science Foundation of China
under Grants Numbers 11765012, 12075126, 11521505, 11621131001and 12105148 and National Key Research and Development Program of China under Contract No.2020YFA0406400.


\end{document}